# A Pseudo-random Number Generator for Multi-Sequence Generation with Programmable Statistics

Jianan Wu, Ahmet Yusuf Salim, Eslam Elmitwalli, Selçuk Köse and Zeljko Ignjatovic

Department of Electrical and Computer Engineering, University of Rochester, Rochester, NY, USA

{jwu144, asalim, eelmitwa}@ur.rochester.edu, selcuk.kose@rochester.edu, zignjato@ur.rochester.edu

*Abstract*—Pseudo-random number generators (PRNGs) are essential in a wide range of applications, from cryptography to statistical simulations and optimization algorithms. While uniform randomness is crucial for security-critical areas like cryptography, many domains, such as simulated annealing and CMOS-based Ising Machines, benefit from controlled or non-uniform randomness to enhance solution exploration and optimize performance. This paper presents a hardware PRNG that can simultaneously generate multiple uncorrelated sequences with programmable statistics tailored to specific application needs. Designed in 65nm process, the PRNG occupies an area of approximately 0.0013mm² and has an energy consumption of 0.57pJ/bit. Simulations confirm the PRNG's effectiveness in modulating the statistical distribution while demonstrating high-quality randomness properties.

*Keywords—pseudo-random number generator, programable statistics, multi-sequence generation, CMOS hardware implementation*

## I. Introduction

Random number generation is a fundamental concept in the field of computer science and data analysis. In general, random number generators (RNGs) are categorized into two main types: True Random Number Generators (TRNGs) and Pseudo-Random Number Generators (PRNGs). TRNGs derive randomness from physical phenomena, such as atmospheric noise, thermal noise, or radioactive decay [1]. In contrast, PRNGs utilize deterministic algorithms to generate sequences of numbers that approximate the properties of random sequences. This characteristic, along with their cost-effectiveness and simplicity of implementation, makes PRNGs highly appealing for many uses. Some widely used PRNGs include the Linear Congruential Generator (LCG), Linear feedback shift register (LFSR), Cellular Automata (CA) and Chaotic PRNG [2].

When evaluating the performance of PRNGs, several key metrics are considered, including uniformity, sample independence, large period, reproducibility, consistency, and others, [3]. Among these, uniformity is one of the most critical metrics. It ensures that each bit, whether '0' or '1', has an equal probability of being generated. This is vital in cryptography and statistical simulations, where a lack of uniformity can introduce biases that compromise security or lead to inaccurate results. However, not all applications require or benefit from perfectly uniform distribution. In certain cases, like simulated annealing or Bistable Resistively-coupled Ising Machines (BRIM) [4][5], a controlled non-uniform distribution can be more advantageous. These applications often require multiple uncorrelated sequences with adaptive randomness to enhance solution space exploration or optimize algorithmic performance. This paper presents a hardware PRNG that allows precise control over the output statistical randomness and enables the simultaneous generation of various sequences. The remainder of this paper is organized as follows: Section II details the design and hardware implementation of the proposed PRNG. Section III presents the simulation results and analysis, while Section IV concludes the paper with a summary of key findings.

## II. PRNG with Programmable Statistical Randomness

The proposed PRNG design integrates an LFSR to generate uniformly distributed pseudo-random samples, a flexible threshold controller for modulating output statistics, and a digital comparator for producing the final pseudo-random sequence, as shown in Fig. 1. By selecting *m* sets of taps from the LFSR, each containing several taps, these sets can be fed into XOR gates to produce an *m*-bit pseudo-random sequence. The threshold controller outputs an *m*-bit number, which is then compared against the generated *m*-bit pseudo-random sequence by an *m*-bit digital comparator, generating a 1-bit sequence with programmable statistical properties. This programmability enables the tuning of the statistics of the 1-bit output sequence by adjusting the threshold controller's output.

When multiple independent 1-bit sequences are required, the design can be easily extended by adding one comparator and the corresponding *m* XOR gates for each additional sequence, while sharing the LFSR and threshold controller across all sequences. This approach enhances power and area efficiency, making it ideal for applications requiring multiple pseudo-random sequences simultaneously, such as Ising machines in [4] and [5].

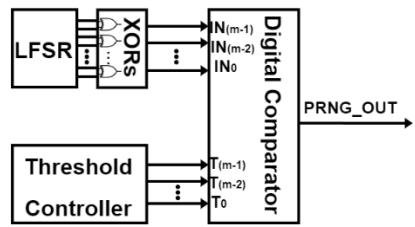

Fig. 1. Block diagram of the proposed PRNG.

### A. Linear Feedback Shift Register

LFSR is a widely used structure in digital circuits that can generate pseudo-random sequences with long periods. The LFSR consists of a series of flip-flops connected in sequence, along with a feedback path that gathers signals from certain flip-flop outputs and combines them using XOR operations before feeding the result back to the input of the leftmost flip-flop. The behavior of an LFSR is typically described by a characteristic polynomial that defines its feedback structure. The general form of the characteristic polynomial is:

$$P(x) = x^n + a_{n-1}x^{n-1} + \cdots + a_1 x + 1 \quad (1)$$

where $x^i$ represents the $i^{th}$ flip-flop, $a_i$ is either 0 or 1, indicating whether the $i^{th}$ flip-flop participates in the feedback signal. For example, a 3-bit LFSR [6] shown in Fig. 2. can be represented by the polynomial $P(x) = x^3 + x^2 + 1$, where the outputs of the 2$^{nd}$ and 3$^{rd}$ flip-flops are XORed to generate the feedback signal.

The length of the sequence generated by the LFSR depends on its characteristic polynomial. If the characteristic polynomial is irreducible over $\mathbb{F}_2$ [3], it is called a primitive polynomial. An LFSR that uses a primitive polynomial can generate a maximum-length sequence, with a period of $(2^n - 1)$, where $n$ is the number of flip flops in the LFSR.

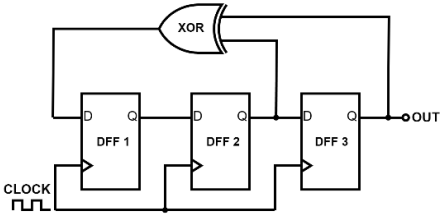

Fig. 2. A 3-bit LFSR architecture.

In order to generate $m$-bit uniformly distributed pseudo-random sequences, careful selection of taps is essential. This is because two sets of taps with the same interval produce two highly correlated sequences through XOR operations; in fact, one sequence can be obtained by shifting the other. To ensure that the pseudo-random sequences generated by the XOR operations are uncorrelated, taps with different intervals must be chosen for each XOR gate. For example, when only two taps are XORed, the number of unique intervals is limited to ($n$-1), where $n$ is the number of bits in the LFSR. Therefore, the number of uncorrelated $m$-bit sequences that can be generated is restricted to ($n$-1)/$m$. This becomes problematic when many uncorrelated pseudo-random sequences need to be generated. To address this issue, multiple taps can be XORed together. By increasing the number of taps in each XOR operation, the potential number of uncorrelated $m$-bit sequences increases significantly, with the total number being $C(n$-1, $k$-1)/$m$, where $k$ is the number of taps in each XOR operation.

*B. Threshold Controller*

The core function of the threshold controller is to adjust the output of the digital comparator by modulating its threshold, thus controlling the final statistical distribution of the pseudo-random sequence. Due to the $m$-bit pseudo-random sequence generated by the combination of LFSR and XOR gates being uniformly distributed, the statistical properties of the output sequence depend solely on the output of the threshold controller. By changing the threshold, different statistical distributions of the pseudo-random sequences can be generated. For example, the threshold controller could be used to adjust the threshold over time enabling the system to start the annealing process in CMOS-based Ising machines with a higher degree of uniformity and then gradually reduce it over time to aid in machine's convergence towards optimal solutions. An example dynamic threshold controller implementation is illustrated in Fig. 3. The counter and logic circuit collaboratively adjust the threshold dynamically. The counter's output serves as the threshold controller's output, and its initial value can be flexibly set to modify the initial statistical distribution. As the counter value increases over time, when all bits reach '1', the inverter output turns '0', cutting off the clock signal, causing the counter to stop counting and hold at its maximum value.

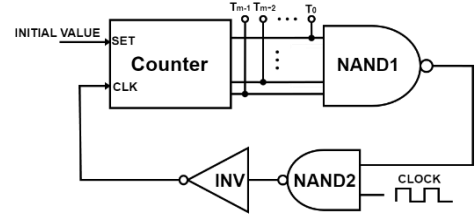

Fig. 3. A counter-based dynamic threshold controller.

*C. Digital Comparator*

A traditional $m$-bit digital comparator is used to compare two $m$-bit binary numbers, $A$ and $B$, and expresses the comparison result through three output bits. Among these three output bits, only one bit will be set to '1', with the following conditions: when $A > B$, the first output bit is set to '1', indicating that $A$ is greater than $B$, while the other two bits are set to '0'; when $A = B$, the second output bit is set to '1', indicating that $A$ is equal to $B$, and the remaining bits are set to '0'; and when $A < B$, the third output bit is set to '1', indicating that A is less than $B$, with the other two bits set to '0'. This design ensures that at any given time, only one state is set to '1', clearly reflecting the comparison relationship between the two numbers.

In our design, the objective is to compare uniformly distributed $m$-bit pseudo-random sequence against a threshold in order to generate a binary sequence (0's and 1's) whose statistical randomness is dependent on the threshold. To achieve this, the output of the digital comparator has been simplified, retaining only the first output bit (i.e., when $A > B$, the output is '1'; when $A \leq B$, the output is '0'). Here, A is an $m$-bit pseudo-random number generated at the output of the XOR logic block, while $B$ is the $m$-bit output from the threshold controller. This design allows us to effectively utilize the comparator to generate a pseudo-random binary sequence with programmable statistical randomness. The theoretical expression of the probability of generating '1' can be expressed by the following equation, where *Threshold* denotes the output of the threshold controller.

$$P(1) = \frac{(2^m - 1) - Threshold}{2^m} \quad (2)$$

III. SIMULATION RESULTS AND ANALYSIS

The proposed PRNG was designed using 65nm CMOS process technology and simulated in the Cadence Virtuoso Analog Design Environment (ADE). A 32-bit LFSR was chosen to ensure the generated sequence has a sufficiently long period, thereby providing extensive state coverage, along with several XOR gates used to perform XOR operations on different bits of the LFSR. An 8-bit comparator and an 8-bit threshold controller were selected for precise tuning of the statistical distribution of the PRNG output while maintaining low power consumption. The layout is shown in Fig. 4., with an area of approximately 61.5μm×21.2μm=0.0013mm$^2$. At a clock rate of 2GHz, the power consumption is around 1.14mW, resulting in an energy consumption of 0.57pJ/bit.

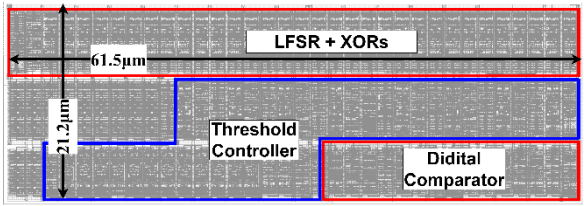

Fig. 4. Layout of the proposed PRNG in 65nm CMOS process.

The remainder of this section presents the simulation results along with a detailed analysis of the PRNG's performance. First, the programmability of statistical randomness in the proposed PRNG is evaluated by adjusting the threshold to analyze the output sequence's statistical distribution. Second, the cross-correlation and auto-correlation of the generated sequences are investigated to assess the quality of their randomness.

A. *Programmability Evaluation*

Fig. 5. illustrates sample output sequences generated at thresholds of 27, 127, and 227, to clearly demonstrate how thresholds influence the statistical randomness of the proposed PRNG. These examples clearly demonstrate that as the threshold increases, the probability of generating 1's in the output sequence decreases, highlighting the significant influence of threshold settings on the statistical behavior of the PRNG.

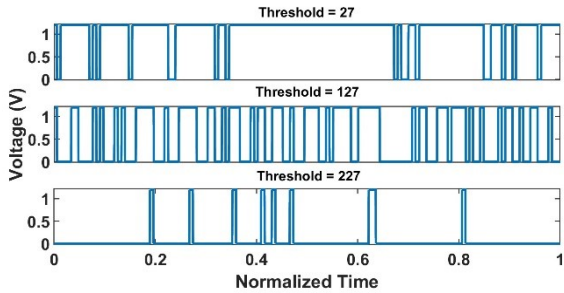

Fig. 5. The output sequences generated at different fixed thresholds.

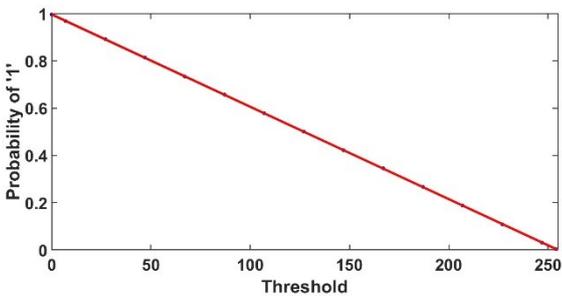

Fig. 6. Threshold-dependent variation in probability of '1'.

Following this, Fig. 6. provides a quantitative analysis by plotting the probability of generating 1's as a function of the threshold, with results closely matching the theoretical expectation in Eq. (2) for *m*=8. The observed pattern in the graph confirms the effectiveness of the design in achieving programmable statistical randomness.

Building upon the previous analysis of fixed threshold control, the next step is to examine how the PRNG behaves under a dynamic threshold controller. The dynamic threshold controller continuously adjusts the threshold over time, which allows the PRNG to exhibit changing statistical properties as the system evolves. Fig. 7. illustrates the performance of the PRNG when using a counter-based dynamic threshold controller shown in Fig. 3. In the figure, the threshold is initially set to '0', and as the dynamic threshold controller operates, the threshold increases over time until it reaches its maximum value. The blue points in the graph represent the measured normalized cumulative count of 1's as it evolves over time. This process can be understood in two stages: before and after the threshold reaches its maximum.

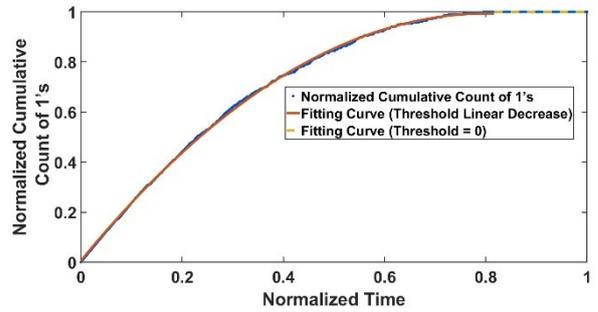

Fig. 7. Cumulative count of 1's over time with dynamic threshold control.

Before the threshold reaches its maximum, the normalized cumulative count of 1's exhibits a quadratic trend, which is depicted by the red curve and can be fitted with the equation

$$CC_N(1) = -1.5396 t_N^2 + 2.4658 t_N + 0.0055 \qquad (3)$$

where $CC_N$ represents the normalized cumulative count of 1's, and $t_N$ denotes the normalized time. The derivative of this quadratic function represents the rate at which the cumulative count of 1's grows.

$$\frac{dCC_N}{dt_N} = -3.0792 t_N + 2.4658 \qquad (4)$$

This rate of growth decreases linearly as time progresses, reflecting the fact that as the threshold increases, the probability of outputting 1's decreases. This behavior aligns with the design, where the threshold increases linearly over time, leading to a corresponding linear decrease in the likelihood of generating 1's. Once the threshold reaches its maximum value, the cumulative count stabilizes, as shown by the yellow line representing $CC_N = 1$. At this stage, no further 1's are generated, and the output remains zero.

B. *Randomness Quality Evaluation*

Cross-correlation assesses the similarity between two different sequences *x(n)* and *y(n)*. It can be defined as:

$$R_{xy}(f) = \frac{\sum_{n=1}^{n=N}(x(n)-\bar{x})(y(n+f)-\bar{y})}{\sqrt{\sum_{n=1}^{n=N}(x(n)-\bar{x})^2} \cdot \sqrt{\sum_{n=1}^{n=N}(y(n)-\bar{y})^2}} \quad (5)$$

where $N$ is the length of the sequences, $\bar{x}$ and $\bar{y}$ are the means of the sequences $x(n)$ and $y(n)$, respectively, and $f$ is the lag, indicating the time delay or shift of one sequence relative to the other. It is essential that both $n$ and $n+f$ in the numerator remain within the valid index range, i.e., $1 \leq n \leq N$ and $1 \leq n+f \leq N$.

Several sequences were generated and their cross-correlation was calculated at various threshold values. Fig. 8. shows the maximum cross-correlation values for sequences generated at different thresholds. The maximum cross-correlation represents the strongest dependency between the two sequences at any lag, defined as either the highest positive or negative correlation value from the cross-correlation function, depending on which has the greater absolute value. As shown in Fig. 8., the maximum cross-correlation values are nearly zero, indicating that the sequences with similar statistical randomness do not exhibit significant cross-correlations and are largely independent.

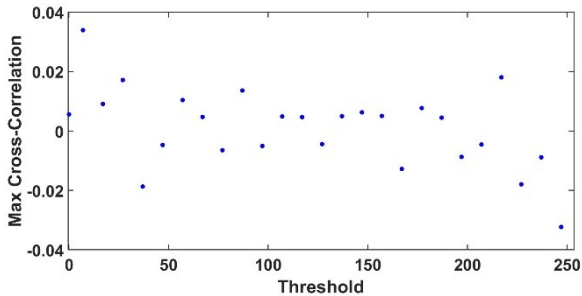

Fig. 8. Maximum cross-correlation at different thresholds.

Auto-correlation, which evaluates the randomness of generated sequences, shares a similar formula with cross-correlation. It effectively calculates the correlation of a sequence with itself. Thus, the auto-correlation can be expressed as:

$$R_{xx}(f) = \frac{\sum_{n=1}^{n=N}(x(n)-\bar{x})(x(n+f)-\bar{x})}{\sum_{n=1}^{n=N}(x(n)-\bar{x})^2} \quad (6)$$

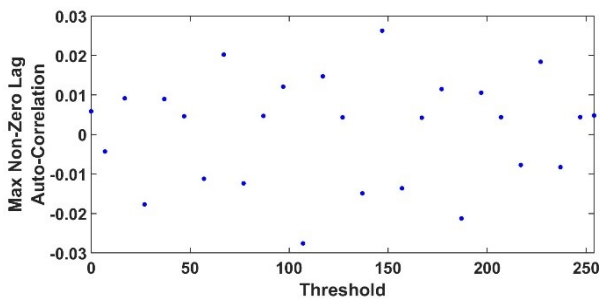

Fig. 9. Maximum auto-correlation at non-zero lags for different thresholds.

The formula shows that the auto-correlation at lag 0 is always equal to 1. For non-zero lags, when the auto-correlation is close to 0, it reflects good randomness, with minimal predictability. In contrast, a significantly large auto-correlation at non-zero lags suggests periodicity or repeating patterns in the sequence. Fig. 9. illustrates the maximum auto-correlation values at non-zero lags for sequences generated at various thresholds. As shown, these values are close to zero, indicating that the generated sequences exhibit strong randomness.

IV. CONCLUSIONS

The design and analysis of a PRNG for multi-sequence generation with programmable statistical randomness, implemented using 65nm CMOS process technology, have been presented. This design allows for precise control of the output sequence distribution through threshold controllers, enabling tunable randomness. Additionally, the number of output sequences can be easily expanded by adding digital comparators and XOR gates, while sharing the same LFSR and threshold controller, offering scalability without significantly increasing hardware overhead. The design occupies an area of about 0.0013mm², with a power consumption of approximately 1.14mW at a clock rate of 2GHz, leading to an energy consumption of 0.57pJ/bit. Simulations were conducted to evaluate the PRNG's performance, with results closely aligning with theoretical expectations, thereby demonstrating the effectiveness of threshold-based control in modulating the statistical properties of the generated pseudo-random sequences. Furthermore, the cross-correlation and auto-correlation of the generated sequences were analyzed. The near-zero cross-correlation, along with the low auto-correlation at non-zero lags, confirmed the strong randomness in the sequences. The proposed PRNG design strikes a balance between flexibility and performance, making it suitable for applications that require multiple sequences with customizable randomness.

ACKNOWLEDGMENT

This work was supported in part by NSF under Award No. 2233378 and by DARPA under contract No. FA8650-23-C-7312.